\documentclass[twocolumn,superscriptaddress,showpacs,prd,aps,amsmath,amssymb]{revtex4-1}

\usepackage{graphicx,psfrag}
\usepackage{color}
\usepackage{amsmath,amsfonts,amssymb,amsthm}
\usepackage{hyperref}
\usepackage{times}
\usepackage{color}

\usepackage{braket}
\usepackage{natbib}
\usepackage{verbatim}
\usepackage{hyperref}
\usepackage{float}
\usepackage{wasysym}
\usepackage{amssymb,graphicx}
\usepackage{epsfig}
\usepackage{psfrag}
\usepackage{dsfont}
\usepackage{amsfonts}
\usepackage{mathrsfs}
\usepackage{multirow}
\usepackage{times}
\usepackage{bm}
\usepackage{hyperref}
\begin{document}
\title{ Magnetic Braking and Damping of Differential Rotation in Massive Stars}
\author{Lunan Sun}
\affiliation{Department of Physics, University of Illinois at
  Urbana-Champaign, Urbana, IL 61801}
\author{Milton Ruiz}
\affiliation{Department of Physics, University of Illinois at
  Urbana-Champaign, Urbana, IL 61801}
\author{Stuart L. Shapiro}
\affiliation{Department of Physics, University of Illinois at
    Urbana-Champaign, Urbana, IL 61801}
\affiliation{Department of Astronomy \& NCSA, University of
  Illinois at Urbana-Champaign, Urbana, IL 61801}

\begin{abstract}
  Fragmentation of highly differentially rotating massive stars that undergo collapse
  has been suggested as a possible channel for binary black hole formation. Such a
  scenario could explain the formation of the new population of massive black holes
  detected by the LIGO/VIRGO gravitational wave laser interferometers. We probe that
  scenario by performing general relativistic magnetohydrodynamic simulations of
  differentially rotating massive stars supported by thermal radiation pressure
  plus a gas pressure perturbation. The stars are initially threaded by
  a dynamically weak, poloidal magnetic field confined to the stellar interior. We
  find that magnetic braking and turbulent viscous damping via magnetic
  winding and the magnetorotational instability in the bulk of the star redistribute
  angular momentum, damp differential rotation  and induce the formation of a massive
  and nearly uniformly rotating inner core surrounded by a Keplerian envelope.  The
  core + disk configuration evolves on a secular timescale and remains in quasi-stationary
  equilibrium until the termination of our simulations. Our results suggest that the high
  degree of differential rotation required for $m=2$ seed density perturbations to trigger
  gas fragmentation and binary black hole formation is likely to be suppressed during the normal
  lifetime of the star prior to evolving to the point of dynamical instability to collapse.
  Other cataclysmic events, such as stellar mergers leading to collapse, may therefore be
  necessary to reestablish sufficient differential rotation and density perturbations
  to drive nonaxisymmetric modes leading to binary black hole formation.
\end{abstract}
\maketitle
\bigskip
%
\section{Introduction}
\label{section:Introduction}
The multiple detections by the LIGO/VIRGO Scientific Collaboration (LVSC) of
gravitational wave (GW) signals produced by binary black hole (BH) mergers
\cite{LIGO_first_direct_GW,Abbott:2016nmj, Abbott:2017,Abbott:2017oio,GW170608,LIGOScientific:2018mvr}
provides the clearest evidence of the existence and common occurrence of close
binary BHs. These detections also point to a new population  of stellar-mass
binary BHs whose masses are larger than those of the twenty X-ray BH binaries with
masses $\lesssim 20M_\odot$~\cite{RemMcC06}, and it raises a puzzle regarding
the origin of these massive BH binaries.

There are various scenarios proposed to explain the formation of massive BH binaries,
such as the coalescence of primordial black holes~\cite{BirChoMun16, CarKuhSan16,
  CleGar17, Ero18} and failed supernova explosion~\cite{SchBatRam18}. 
The collapse of a massive star ($M \gtrsim 10M_\odot$) has been suggested as one of
the most plausible formation channels for BHs observed by the LVSC~\cite{TutYun93,
  VosTau03, SchBatRam18}. As the final mass of the BH remnant  depends sensitively
on the mass and metallicity of the progenitor stars~\cite{WooHegWea02}, it is
somehow expected that the direct--collapse of Population III (Pop III) stars form the
most massive BHs~\cite{BelBulFry10,DomBelFry13}. This hypothesis was confirmed
by population synthesis simulations of Pop III stars~\cite{KinInaHot14}, where
binary BHs with typical mass of~$\sim 30 M_{\odot}$ were formed from $10^6$ Pop III
binary evolutions with masses ranging between $10M_\odot$ to  $100 M_\odot$.
The study in~\cite{BelHolBui16} suggests that a binary containing massive stars with
masses $M_1 \approx 96 M_{\odot}$ and  $M_2 \approx 60M_{\odot}$ at redshift $z\approx
3.2$ may be the progenitor of the first BH binary detected by the LVSC (event GW150914)
\cite{LIGO_first_direct_GW}. Later, it was shown in~\cite{SteVigMan17} that
the events GW150914 and GW151226, as well as the lower-confidence event LVT151012,
may be explained by the isolated binary evolution channel (i.e. assuming that the binary
never interacts with another star) from
progenitor stellar binaries with masses between $60M_\odot-160M_\odot$. This channel
involves the formation of a common envelope through which  mass is transferred between
the secondary star and its companion via stellar winds. However, such a common-envelope
phase is not well understood and results in great physical uncertainties for the BH
formation rate due to the lack of observational evidence ~\cite{KalKimLor04}. To avoid
this phase, an evolution channel including massive overcontact binaries, where two tightly
bound stars are sufficiently mixed due to their tidally induced high spin, has been
considered~\cite{MarLanPod16}. Prior to collapse, the massive close binary components
are fully mixed and achieve chemical homogeneity. As
the hydrogen burning ends nearly simultaneously for the two binary components, the
post-main-sequence expansion is suppressed due to the absence of a hydrogen-rich
envelope. According to the mass exchange mechanism, this evolution
route is therefore likely to produce BH binaries with a large mass ratio, as in
GW150914~\cite{MinMan16}. 
Recently, a highly spinning and aligned BHBH event in the first observing run (O1) of Advanced
LIGO~\cite{LIGOScientific:2018mvr} was recently reported~\cite{Zackay:2019tzo},  which lends
support to the hypothesis that stellar binaries may be the progenitors of at least some LVSC
BHBHs.

It also has been suggested that stellar-massive BHs can be formed via dynamical interactions
in dense clusters. In clusters with low metallicities, such as the cores of globular clusters
(GCs) or dense AGNs~\footnote{However, low-metallicity AGN is extremely rare from observations.
For details, see, e.g., \url{http://dx.doi.org/10.1111/j.1365-2966.2006.10812.x}.} it
is possible to form binary systems dynamically via gravitational interactions among a population
of stellar-mass BHs~\cite{RodMorLon16, ChaRodRas17} or hierarchical mergers of smaller
BHs~\cite{FisHolFar17}. In particular, a simulation of GCs of mass $M \sim 10^5-
10^6 M_\odot$ with many-body dynamics and stellar evolution has provided an evolutionary channel
for BH binaries similar to GW150914~\cite{RodHasCha16}. It is expected that the two mechanisms
described above can be distinguished by the binary spins from a large, future sample of BH
mergers. The BH binaries formed from direct collapse of massive stars are more likely to result
in aligned spins due to tidal interactions and/or mass transfer, while dynamically formed binaries
will have randomly oriented spins. In addition, the further merging of the massive BH remnants
in GCs may finally form more massive intermediate-mass black holes (IMBHs),
defined as BHs with mass $10^2-10^5M_\odot$~\cite{MilCol03}. An all-sky survey of GW signals
from binary IMBHs has been performed by LVSC. For such systems, the merger rate is constrained
to be~$\gtrsim 0.3 \, \rm{GC^{-1} \, Gyr^{-1}}$ in the local Universe with $90\%$
confidence~\cite{LIGO_IMBH17}. Other evolution channels, such as the direct collapse of very
massive Pop III stars, can also form IMBHs~\cite{MilCol03, GaiManMil11}.

Among the binary BH detections by LVSC, GW150914 and GW170104 are possibly associated with
electromagnetic (EM) counterparts~\cite{ConBurGold16,VerTavUrs17}. Although the EM
emission is unexpected for binary BHs mergers in pure vacuum (see~e.g.~\cite{Lyu16}),
in these two events, gamma-ray signals with isotropic energy $\sim 10^{49}-10^{50} \rm erg$
have been observed shortly after the peak GW signals~\cite{ConBurGold16,VerTavUrs17}.
With the assumption that the EM transient is connected to the GW events, 
fully general relativistic magnetohydrodynamic (GRMHD) simulations
in~\cite{KhaPasRui18} show that magnetized
disk accretion onto a BH binary
could explain GW and EM counterparts such as the GBM hard X-ray signal reported 0.4s after
GW150914. Massive star fragmentation during gravitational collapse has been proposed as
an alternative scenario~\cite{Loe16}. However, numerical simulations and analytic studies
have shown that binary BHs coalescence within a dense stellar gas ($\rho \gtrsim 10^7\rm
g \,cm^{-3} $) produces a waveform different from one in vacuum due to gas drag~\cite{FedOttSpe17,
  DaiMcKMil17}. To rectify this effect, the single-progenitor scenario with an accretion
rate $\sim 3\times10^9$ times the Eddington rate prior to merger has been suggested~\cite{DOrazio:2017bld}.
This so-called giga-Eddington accretion can drive out the
stellar gas as well as alter the time delay between GW and EM signals. However, that
scenario may not account for all the physical processes involved, in particular
magnetically-driven processes~(see e.g.~\cite{Gold:2013zma,Gold2014,KhaPasRui18}), which
can have a dynamical impact on the evolution channel.
%
\begin{table*}[!ht]
\centering
\caption{\label{tab:initial} Initial model parameters.
  Here $n$ denotes the polytropic index,  $M$ is the characteristic mass, $\bar{M}_{\rm ADM}$ is the ADM mass,
  $\bar{\rho}_c$ is the rest-mass central density,  $R_{pol}$ is the polar radius, $\bar{J}_{\rm ADM}$ is
  the ADM angular momentum, ${\mathcal{M}}/|W|$ denotes the ratio of magnetic energy to gravitational potential
  energy, and $T/|W|$ the ratio of rotational kinetic energy to gravitational potential energy. {In
  these cases we have set $\bar{\mathcal{M}}=3.711\times 10^{-6}$}.
  Nondimensional quantities have been rescaled with the polytropic gas constant $K$ and are denoted with a bar.}
\begin{ruledtabular}
\begin{tabular}{ccccccccccc}
Case& $n$ & $M/M_{\odot}$ &$\bar{M}_{ADM}^{(a)}$&$\bar{\rho}_c^{(b)}$ & $R_{pol}/M_{ADM}$ & $\bar{J}_{ADM}^{(c)}$ & $\mathcal{M}{^{(d)}} /|W|$& $T/|W|$ \\
\hline
\textbf{N29}         &  2.9 &$\sim 10^4$&4.139 & $4\times 10^{-8}$ & 161    &36.92	&$1.88\times 10^{-4}$ &0.091\\
\textbf{N295}        &  2.95 &$\sim 10^5$&4.801 & $2\times 10^{-8}$&192 &52.44	&$1.82\times 10^{-4}$ &0.085\\ 
\end{tabular}
\end{ruledtabular}
\begin{flushleft}
  $^{(a)}$  $\bar{M}_{ADM} = K^{-n/2}M_{ADM}$ where $K=P/\rho_0^\Gamma$.\\
  $^{(b)}$  $\bar{\rho}_c = K^{n}\rho_c$.\\
  $^{(c)}$    $\bar{J}_{\rm ADM} =K^{n/2}J_{\rm ADM}$.\\
  $^{(d)}$    $\bar{\mathcal{M}} =K^{-n}\mathcal{M}$.
\end{flushleft}
\end{table*}

Returning to the collapse of massive stars as the origin of massive BHs and massive BH binaries, we
note that massive stars are radiation dominated and well-approximated by an $n\sim 3$
($\Gamma \sim 4/3$) polytrope. During their evolution, very massive (and supermassive) stars cool and
contract until they ultimately encounter a relativistic, dynamical instability leading to catastrophic
collapse to BHs (see~\cite{BauSha99, ShiSha02,SaiThoSha02,ZinSteHaw05,SaiHaw09,ShiUchSek16,ButLimBau18}).
Massive stars that are  \textit{uniformly} rotating remain nearly axisymmetric during the
collapse, as shown by post-Newtonian~\cite{SaiThoSha02} and fully general relativistic (GR) simulations
\cite{SunPasRui17,SunRuiSha18}. However, massive stars that are \textit{differentially} rotating with
a sufficiently high ratio of rotational kinetic energy to gravitational potential energy
($T/|W| \gtrsim 0.14$) are secularly unstable to forming $m=2$ bar modes  during the course of their
evolution~\cite{BodOst73,NewSha00,NewSha00a}. In addition, differentially rotating massive stars
with nonaxisymmetric (e.g. bar mode) density perturbations and undergoing collapse can fragment
into self-gravitating, collapsing components that form seed BHs~\cite{ZinSteHaw05, ZinSteHaw06}.
In some cases, this fragmentation may lead to the formation of massive BH binaries~\cite{ReiOttAbd13}.

Even though intensive studies of the evolution channels and the interactions of 
collapsing massive stars have been studied previously, our understanding of the role of the
magnetically--driven instabilities during their stellar evolution prior to collapse remains
poor.

Newtonian simulations of magnetic braking (i.e.~winding) and viscous damping of
an incompressible, differentially rotating infinite cylinder model were performed in~\cite{Sha00}.
It was found that differential rotation could generate toroidal Alfv\'en waves that convert an
appreciable fraction of rotational kinetic energy to magnetic energy and drive the star toward
uniform rotation (``magnetic braking''). The destruction of differential rotation was also demonstrated
later on in a compressible cylinder model~\cite{CooShaSte03} and in relativistic, 
neutron star models, both incompressible~\cite{LiuSha03} and compressible~\cite{Etienne:2006am}. The
role of turbulent viscosity in damping differential rotation
in rapidly spinning, neutron stars obeying a ``stiff" ($\Gamma = 2$) equation of state (EOS) was
demonstrated by simulations in axisymmetry in~\cite{DueLiuSha04}, where the Navier-Stokes equations
with a shear viscosity were solved in full GR. They found that the viscosity drives the neutron stars
to a high density, uniformly rotating core surrounded by a Keplerian disk in which the core could either
collapse to a BH or remain stable depending on its mass and the viscous build-up of thermal pressure.
Subsequently, the role of magnetic fields in driving viscous turbulence via the magneto-rotational instability
(MRI), which damps the differential rotation in a hypermassive neutron star, was demonstrated by GRMHD
simulations in axisymmetry in~\cite{DueLiuSha05b, DueLiuSha06, ShiLiuSha06,Etienne:2006am}. They showed
that the magnetic energy is amplified both by the winding of field lines and MRI, and that the differential
rotation is significantly reduced due to the magnetic braking and MRI-induced turbulent viscous damping.

In this paper, we  explore this braking and damping scenario in weakly magnetized, differentially
rotating massive stars in hydrostatic equilibrium. In particular, we perform GRMHD simulations to
probe the effects of magnetic winding and magnetic-induced turbulent viscosity that can break the
differential rotation in massive stars.  We consider massive stars governed by a soft $n \lesssim 3$
($\Gamma \gtrsim 4/3$) EOS. We first show that, for magnetized stars with characteristic
masses greater than several hundred $M_\odot$, the hierarchy of physical timescales meets the criteria
for the magnetic damping of differential rotation within stellar lifetimes. Then we construct stable
equilibrium stars with the same differentially rotating profiles as in the initial (collapsing)
configurations employed in~\cite{ReiOttAbd13}, where fragmentation of $m=2$ seed density perturbations
into BH binaries occurs. We endow the stars with a dynamically weak, dipole magnetic field confined to
the stellar interior and evolve the resulting magnetized equilibrium configuration in 3+1 dimensions
using our Illinois GRMHD code. We find that well prior to collapse, magnetic braking and turbulent
viscosity induces the
formation of an inner core surrounded by a thick, Keplerian, circumstellar disk after several Alfv\'en
timescales.  As anticipated, magnetic braking, induced by the toroidal winding of the magnetic field,
and viscous damping, induced by magnetic turbulence triggered by the MRI, drive the inner core to
uniform rotation. The new-born core + disk configuration evolves on a secular (viscous) timescale and
remains in quasistationary equilibrium until we terminate the simulations.  Our results suggest that
{\it a stellar fragmentation scenario for a massive star that evolves in isolation is not a plausible
formation  channel for binary BHs}. Other cataclysmic events, such as the merger of two massive stars,
may be necessary to re-establish sufficient differential rotation to drive nonaxisymmetric modes leading
to fragmentation of a dynamically unstable remnant into binaries during collapse. Our results apply for
stars prior to BH formation with characteristic masses greater than several hundred solar masses.
The masses in our simulations are chosen for demonstration purposes and were selected for their 
microphysical simplicity and computational practicality. However, we do expect qualitative similar
outcomes to apply for lower masses relevant to the LIGO/Virgo BH progenitors.

The structure of the remainder of the paper is as follows. In Sec.~\ref{section:Timescales}, we
calculate several key physical timescales and show that the hierarchy of timescales for plausible
initial configurations satisfy conditions inevitably leading to the magnetic destruction of differential
rotation. In Sec.~\ref{section:Initialdata}, we discuss the properties of typical initial configurations  
and describe the initial models used in our simulations. In Sec.~\ref{section:methods}, we
describe the numerical tools we use, the numerical implementation of the initial data and evolution,
and some of the diagnostics employed to verify the reliability of our simulations. We summarize our
results and compare them with previous work in Sec.~\ref{section:Results}. Finally, we offer
conclusions in Sec.~\ref{section:Conclusions}. We adopt geometrized units ($G=c=1$) throughout
the paper except where stated explicitly. Greek indices denote all four spacetime
dimensions, while Latin indices imply spatial parts only.

\section{Timescales}
\label{section:Timescales}
There are four important timescales related to the evolution of magnetized, differentially rotating,
massive stars:
\begin{enumerate}
\item The {\it dynamical timescale} $t_{\rm dyn}$ determines the rate of collapse
  (``free-fall") of the star to a BH after the onset of dynamical radial instability. It is
  also the timescale for a perturbed, stable star to relax to hydrostatic equilibrium. In order to
  guarantee that the star maintains hydrostatic equilibrium while undergoing magnetic braking
  or other secular perturbations, $t_{\rm dyn}$ must be shorter than these other timescales.
\item The {\it Alfv\'en timescale} $t_{\rm A}$ is the winding timescale of a poloidal magnetic
  field in a differentially rotating star~\cite{Sha00}.
  The magnetic braking of differential rotation usually requires several changes in
  the direction of magnetic field rotation, therefore $t_{\rm A}$ is an appreciable fraction
  of the total braking timescale.
\item The {\it viscous timescale} $t_{\rm vis}$ represents the time over which the turbulent
  viscosity, induced by magnetic field instabilities, damps differential rotation.
\item The {\it thermal timescale} $t_{\rm th}$ determines the lifetime of a very massive star,
  which radiates at the Eddington limit. This timescale must be longer than the above timescales
  in order for the magnetic braking and damping of differential rotation 
  to be completed during the lifetime of a stable star.
\end{enumerate}

The above timescales depend on various factors such as the equation of state,
radius, degree of differential rotation, and initial magnetic field strength. In a polytrope,
they can be described approximately by several parameters: polytropic index $n$, typical
density $\rho$,  characteristic mass $M$, compaction $\mathcal{C} \equiv M/R$, where $R$
is the characteristic radius, and the ratio of magnetic energy to gravitational potential
energy $\mathcal{M}/|W|$. To establish the physical hierarchy of timescales, we scale the
parameters above as in our canonical model~\textbf{N29} (see Table ~\ref{tab:initial}). In
addition, we scale our estimates to a characteristic mass of $10^4 M_{\odot}$ following our
analysis of an $n=2.9$~polytrope at the onset of collapse in~\cite{SunRuiSha18}. Since
the ratio of rotational kinetic energy to gravitational potential energy~$T/|W|\ll 1$,
we can use spherical models in making our rough analytic estimates.
So, the above timescales can be estimated as follows:
%
%
\begin{table*}[!ht]
\centering
\caption{\label{tab:initial_2} Initial model parameters. Here
  $\rho_c$ denotes the central rest-mass density, $R_{\rm pol}$ is the polar radius,
  $\Omega_c$ is the central angular velocity, and $\left<B\right>=\sqrt{8\,\pi\mathcal{M}/V_s}$
  denotes the averaged magnetic  field strength, where $\mathcal{M}$ is the magnetic energy
  and $V_s=\int \sqrt{\gamma}\, d^3x$ is the initial proper volume of the star.}
\begin{ruledtabular}
\begin{tabular}{ccccccccccc}
  Case & $\rho_c$  $\left[\left({M}/{10^4M_{\odot}}\right)^{-2}\rm g \,cm^{-3} \right]$ &
         $R_{\rm pol}$ $\left[\left({M}/{10^4M_{\odot}}\right)\rm cm \right]$              &
         $\Omega_c\,[(M/10^4M_\odot)^{-1}\,\rm s^{-1}]$ &
         $\left<B\right>\left[\left({M}/{10^4M_{\odot}}\right)^{-1} \rm G \right]$ &   \\
\hline
\textbf{N29}       &$4.23\times 10^{3}$    &{$2.38\times 10^{11}$} & 0.016 & {$2.68\times 10^9$}\\
\textbf{N295}      &$2.85\times 10^{3}$    &{$2.84\times 10^{11}$} & 0.012 & {$1.90\times 10^9$}\\ 
\end{tabular}
\end{ruledtabular}
\end{table*}
%
%
\paragraph*{}{1. The \it dynamical timescale} is
\begin{eqnarray}
\label{eq:t_dyn}
t_{\rm dyn} &\sim& \braket{\rho}^{-1/2} \sim\left(\frac{M}{R}\right)^{-3/2} M\\
&\sim& 10^2 \left(\frac{M}{10^4 M_{\odot}}\right)\left(\frac{\mathcal{C}}
{6\times10^{-3}}\right)^{-3/2}
\rm s\,.\nonumber 
\end{eqnarray}

\paragraph*{}{2. The {\it Alfv\'en timescale} can be calculated as
\begin{eqnarray}
\label{eq:T_A_1}
t_{\rm A} &\sim& {R}/{v_A}\\
&\sim& 10^4 \left(\frac{\mathcal{M}/|W|}{2\times 10^{-4}}\right)^{-1/2}
\left(\frac{M}{10^4 M_{\odot}}\right)\left(\frac{\mathcal{C}}{6\times10^{-3}}
\right) ^{-3/2} \rm s\,.\nonumber
\end{eqnarray}
where $v_A = B/\sqrt{4\pi \rho}$ is the Alfv\'en speed and $\rho$ is the characteristic density
of conducting plasma (here the stellar gas).  For our canonical model (see~Sec.~\ref{section:Initialdata}),
the Alfv\'en timescale is typically two orders of magnitude longer than the dynamical timescale.
\paragraph*{}{3. The {\it viscous timescale} is 
\begin{equation}
\label{eq:Tvis_1}
t_{vis} \sim \frac{R^2}{\nu} \sim \frac{R}{\alpha_{\rm ss} \,c_s}\,,
\end{equation}
where $\nu = \alpha_{\rm ss}\,H c_s$ is the shear viscosity. Here $c_s = \left(\partial P/\partial
\rho\right)^{1/2}$ denotes the sound speed. We take the dimension of a viscous vortex $H$ to be
comparable to the stellar radius.  The quantity $\alpha_{\rm ss}$ denotes the Shakura--Sunyaev
$\alpha_{\rm ss}$-disk parameter computed as  (see Eq.~26 in~\cite{PenMcKNar10})
\begin{equation}
\alpha_{\rm ss} =\frac{T^{EM}_{\hat{r}\hat{\phi}}}{P+P_{\rm mag}}\sim \frac{T^{EM}_{\hat{r}\hat{\phi}}}{P}\,,
\end{equation}
where $P$ is the matter + radiation pressure, $P_{\rm mag}$ is the magnetic pressure, and $T^{EM}$ is the
electromagnetic stress energy tensor defined as
\begin{equation}
\label{eq:EM_tensor}
T_{\mu\nu}^{EM} = b^2u_{\mu}u_{\nu}+b^2g_{\mu\nu}/2-b_{\mu}b_{\nu}\,,
\end{equation}
where $b^\mu=B^\mu_{(u)}/(4\,\pi)^{1/2}$  is  the magnetic field measured by an observer
co-moving with the fluid, $b^2=b^\mu\,b_\mu$, and $u^\mu$ denotes the
4-velocity of the plasma (see~\cite{BSBook} for discussion). Using~Eq.~\ref{eq:EM_tensor},
and the fact that  $c_s \sim (P/ \rho)^{1/2} \sim (M/R)^{1/2}$, $\alpha_{\rm ss}$ can be estimated as
\begin{equation}
\alpha_{\rm ss} \sim \frac{B^{\hat{r}}B^{\hat{\phi}}}{P}\,.
\end{equation}
Accordingly,
\begin{equation}
  t_{\rm vis} \sim \frac{1}{\alpha_{\rm ss}}\,M^{-1/2}\,R^{3/2}\sim\frac{1}{\alpha_{\rm ss}}\,
  t_{\rm dyn}\,.
\end{equation}
Using~Eq.~\ref{eq:t_dyn} and an effective viscous parameter $\alpha_{\rm ss}~\sim 10^{-4}$,
which is the maximum mean value for $\alpha_{\rm ss}$ during our simulations, we obtain 
\begin{equation}
\label{eq:T_vis}
t_{vis} \sim 10^5\,\left(\frac{\alpha_{\rm ss}}{10^{-4}}\right)^{-1}\,\left(\frac{M}{10^4\,
  M_{\odot}}\right)\,\left(\frac{\mathcal{C}}{6\times10^{-3}}\right)^{-3/2} \rm s\,.     
\end{equation}
In our canonical model, the viscous timescale is  one order of magnitude
longer than the Alfv\'en timescale (see Eq.~\ref{eq:T_A_1}). However, this estimate
is very crude due to the local variations of $\alpha_{\rm ss}$, which may vary by an order
of magnitude at different locations. 
\paragraph*{}{4. The {\it thermal timescale} is defined as
\begin{equation}
\label{eq:T_th_1}
t_{th} \equiv \frac{E_{\rm th}}{L_{\rm Edd}}\,,
\end{equation}
where $E_{\rm th}$ is the thermal energy of the star and $L_{\rm Edd} = 4\pi M/\kappa$ is the
Eddington luminosity with $\kappa$ the opacity~\cite{BauSha99}. The thermal energy
can be estimated from the Newtonian virial theorem
\begin{equation}
2\,T+W+3\,(\Gamma -1)\,E_{th}+\mathcal{M} = 0\,.
\end{equation}
For small $T/|W|$, $\mathcal{M}/|W|$ and $\Gamma\gtrsim 4/3$, the virial theorem gives
\begin{equation}
\label{eq:T_th_2}
W+ E_{th}\approx 0\,.
\end{equation}
Therefore, we obtain $E_{\rm th} \approx |W|\sim M^2/R$, and Eq.~\ref{eq:T_th_1} becomes
\begin{eqnarray}
\label{eq:T_th}
t_{\rm th}&\sim& 10^{12}\left(\frac{\mathcal{C}}{6\times
  10^{-3}}\right) \rm s 
\sim 10^5\left(\frac{\mathcal{C}}{6\times10^{-3}}\right)
\rm yrs\,,
\end{eqnarray}
where we have assumed $\kappa = 0.4 \rm \, cm^2 g^{-1}$, appropriated for Thomson scattering in
a hydrogen gas, and hence $L_{\rm Edd} \sim 10^{42} (M/10^4 M_\odot)\,\rm erg \,s^{-1}$. In our
canonical case \textbf{N29} the thermal timescale $t_{\rm th}$ is therefore about $\sim 10^7$
times larger than  the viscous timescale (see Eq.~\ref{eq:T_vis}). 

To achieve the magnetic damping of differential rotation during the stellar lifetime of an
equilibrium star prior to reaching the onset of dynamical instability to radial collapse, the
evolution of the massive star should satisfy the following hierarchy of timescales
\begin{equation}
\label{eq:hierarchy}
t_{\rm dyn} < t_{\rm A} < t_{\rm vis} (\sim t_{\rm damping}) < t_{\rm th}(\sim
t_{\rm lifetime})\,.
\end{equation}
In particular, the values for case \textbf{N29} in Table~\ref{tab:initial} provide a model which
satisfies this typical hierarchy.
In general, stars with several hundred solar masses are radiation dominated and have $2.9
\lesssim n \lesssim 3$ (see Eq.~\ref{eq:nvsmass} below). Hence, the hierarchy of timescales
in Eq.~\ref{eq:hierarchy} applies to such  stars and strongly suggests that differential rotation
will be damped prior to BH formation. This result is demonstrated by our detailed simulation below.
%
\begin{figure*}[t]
\includegraphics[scale=0.131]{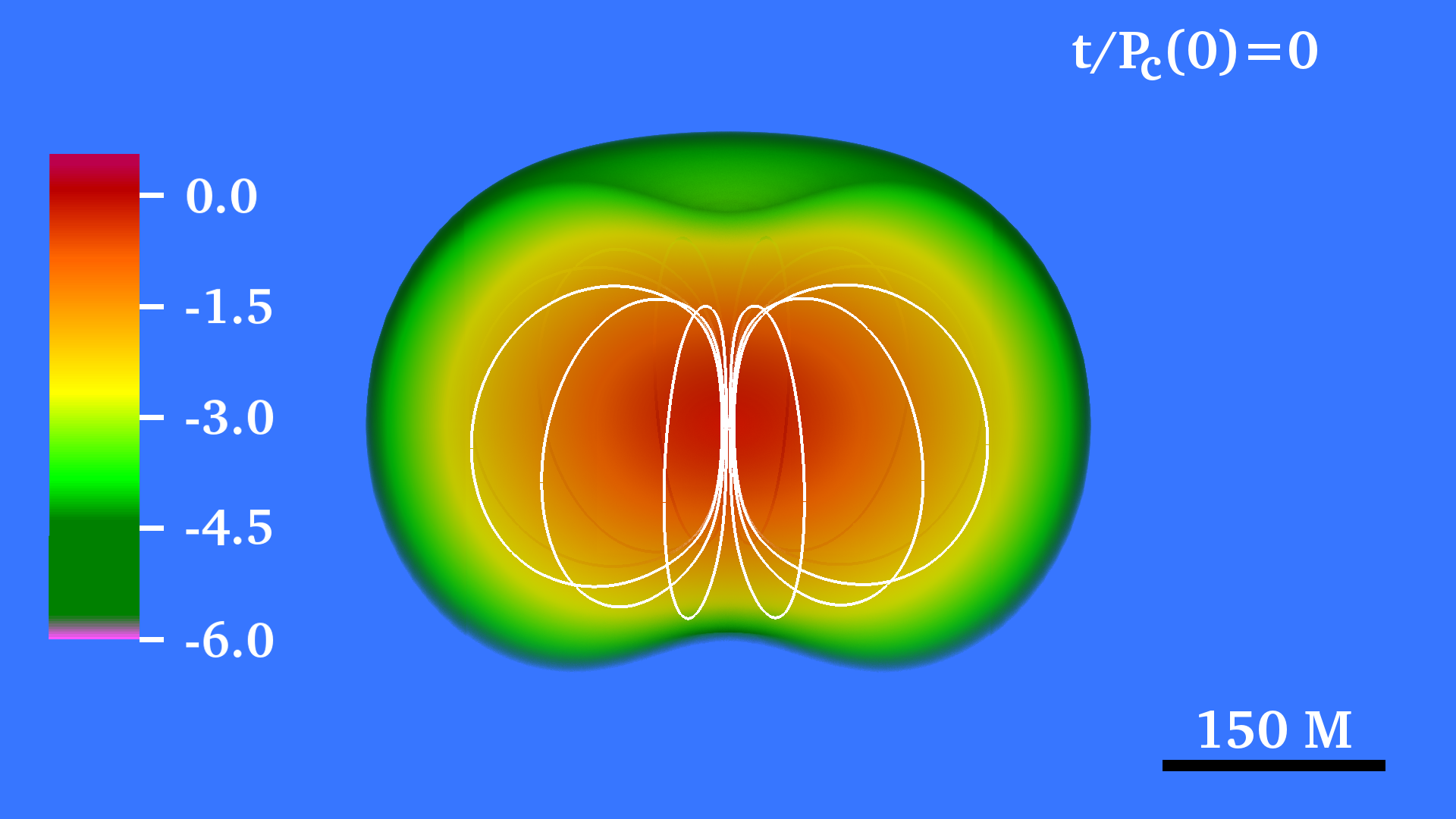}
\includegraphics[scale=0.131]{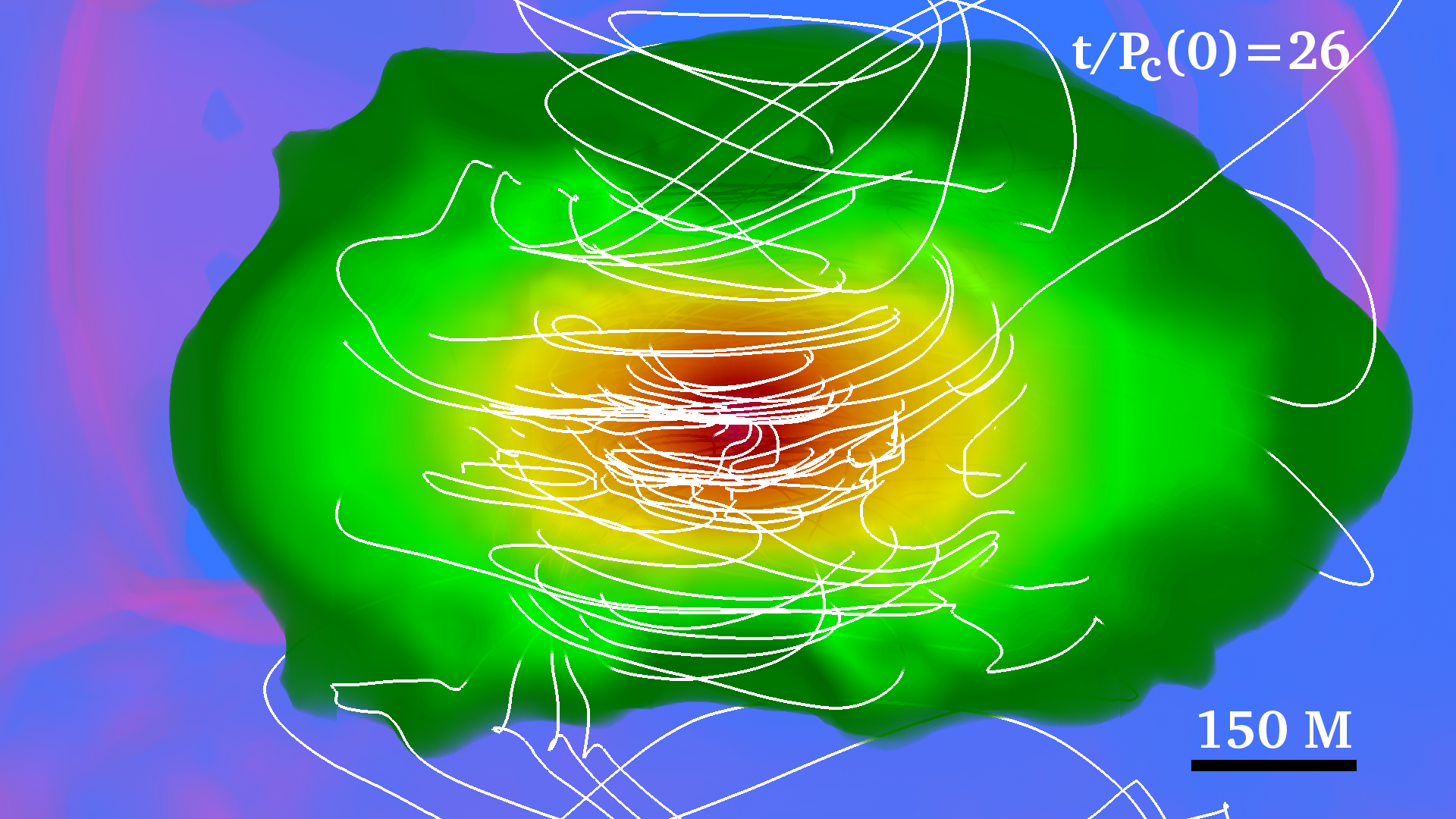}
\includegraphics[scale=0.131]{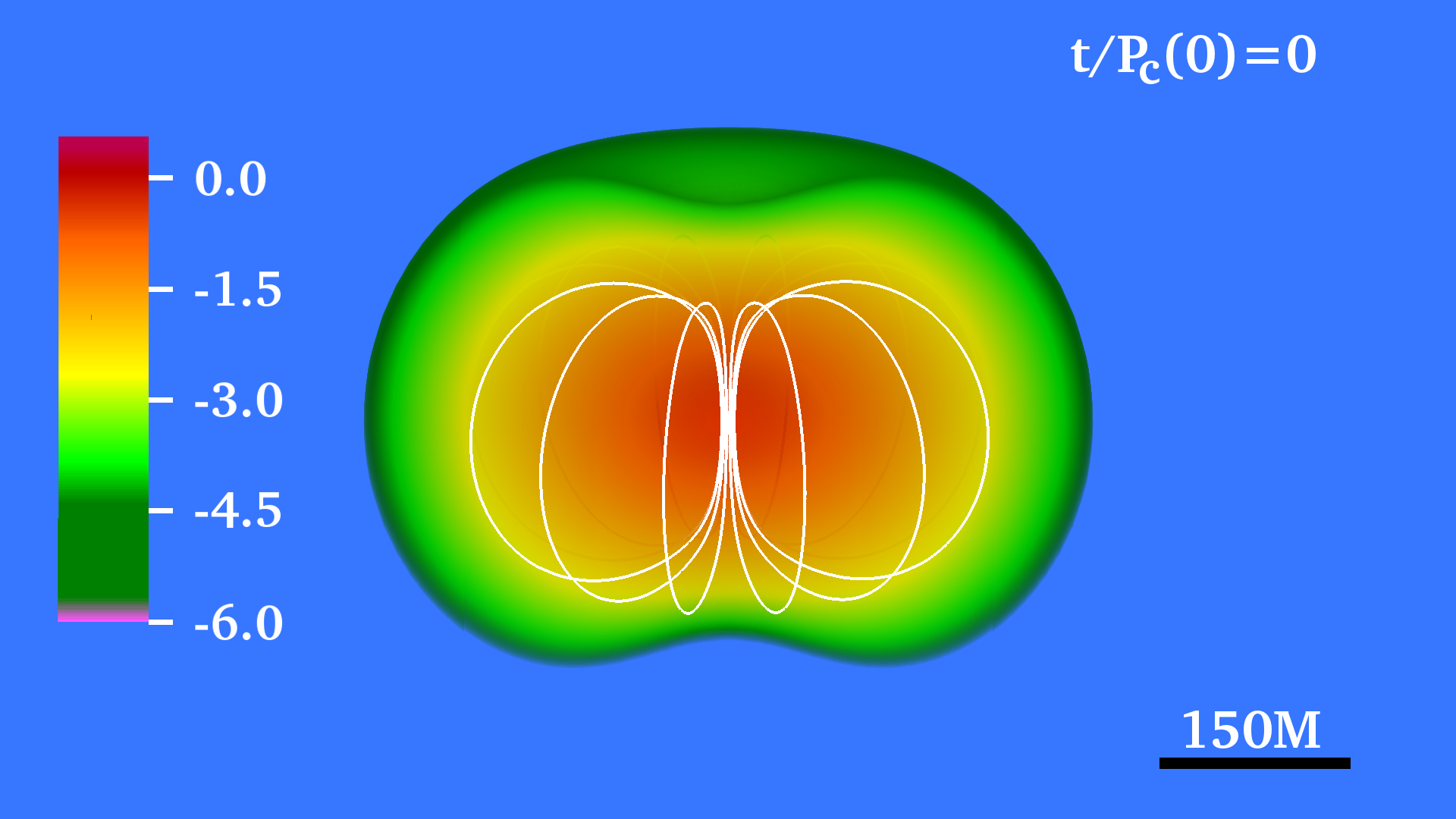}
\includegraphics[scale=0.131]{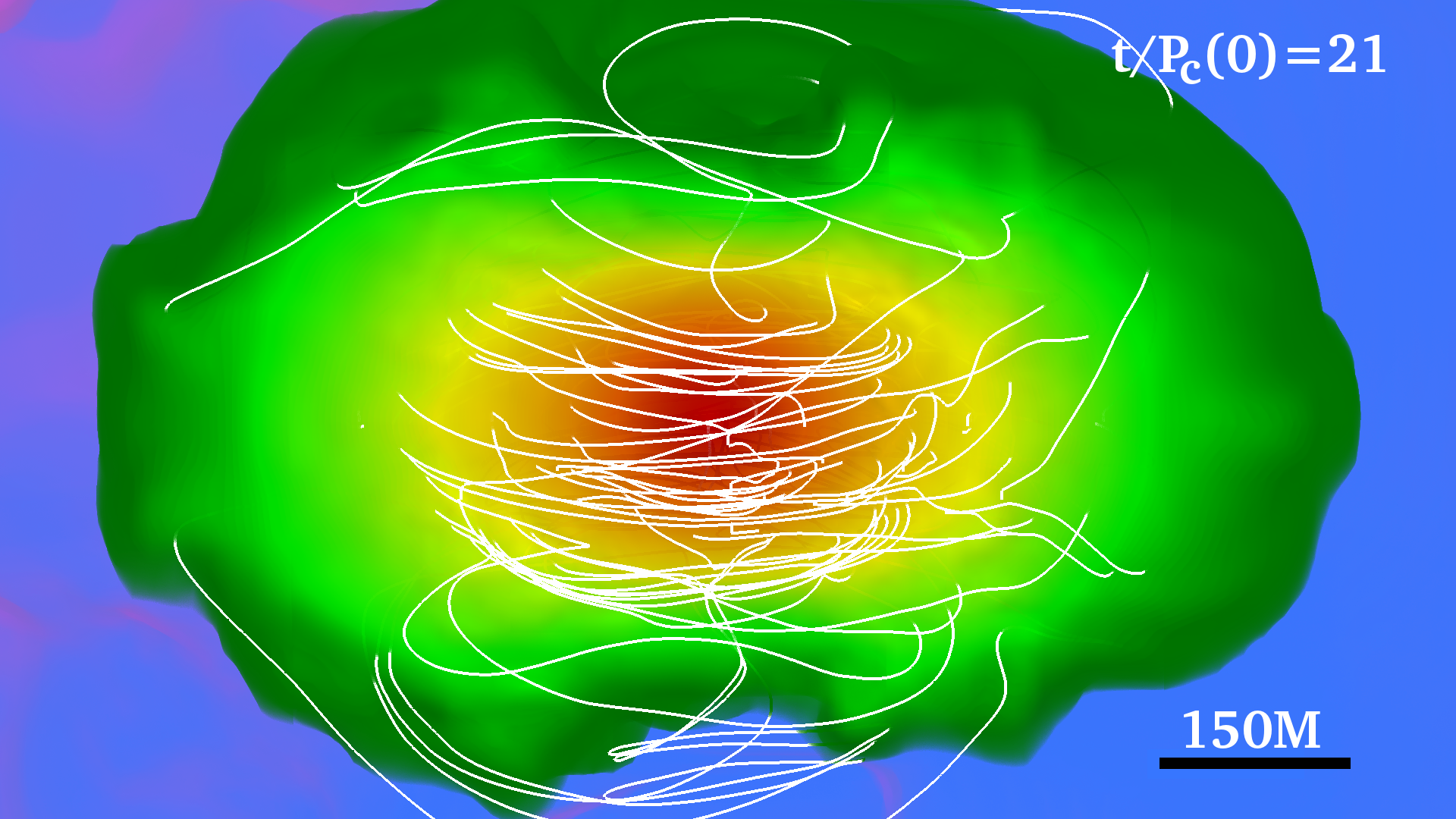}
\caption{\label{fig:density_B} Meridional cut through a 3D volume rendering  of the rest-mass
  density, normalized to its initial maximum value $\rho_{0,max}$  (log scale) for initial (left
  panels) and  final configurations (right panels) for  cases \textbf{N29} (top row) and \textbf{N295}
  (bottom row). For details see Table~\ref{tab:initial}. White solid lines denote the magnetic
  field lines. Here $P_{\rm c}$ denotes the initial central rotation period, which is
  $P_{\rm c}= 8193M$ for \textbf{N29} and $P_{c} = 10582M$ for $\textbf{N29}$, with $M=0.049(M/10^4
  M_\odot){\rm s}=1.47\times 10^4(M/10^4M_\odot)\rm km$.}
\end{figure*}
%
\section{Initial Data}
\label{section:Initialdata}
To numerically study magnetic braking and damping of differential rotation in massive stars,
we consider initial data satisfying the following criteria:
\begin{enumerate}
\item An EOS dominated by thermal radiation pressure perturbed by thermal gas pressure.
  We consider a polytropic EOS with $n=2.9$ and $n=2.95$ (see Table~\ref{tab:initial}).
\item Stars in hydrostatic equilibrium with a lifetime longer than the viscosity timescale
  (see Eq.~\ref{eq:T_vis}).
\item A small magnetic field to induce magnetic winding and an effective turbulent viscosity.
  The stars are seeded with a dynamically unimportant poloidal magnetic field ($\mathcal{M}/
  |W|\sim 10^{-4}$ (See Table~\ref{tab:initial}). 
\item A damping timescale realizable given our computational resources, which requires a
  significant initial magnetic field, albeit one that is dynamically weak with $\mathcal{M}/|W|
  \ll 1$. We aim to provide 
  an ``existence proof'' for the magnetic braking and damping of differential rotation in massive
  stars with the finite computational resources at our disposal.
\end{enumerate}

A radiation-dominated stellar model with radiation pressure $P_{\rm rad}$
and a small perturbation due to gas pressure $P_{\rm gas}$ can be described by a polytrope
with polytropic index $n$ slightly less than 3 (or equivalently, with adiabatic index $\Gamma$
slightly greater than $4/3$). In particular, the fraction of gas pressure $P_{\rm gas}$ is 
quantified by the parameter
\begin{equation}
\beta =  P_{\rm gas}/P_{\rm rad}\,.
\end{equation}
The relation between $n$ ($\Gamma$) and the characteristic mass $M$ is given by
\begin{equation}
\label{eq:nvsmass}
n \approx \frac{3}{1+4.242\left({M}/{M_{\odot}}\right)^{-1/2}}\,,
\end{equation}
which strictly holds for nonrotating, spherical stars. 
We build rotating equilibrium stellar models using the GR equilibrium code described
in~\cite{CooShaTeu92,CooShaTeu94,Cook1994ApJ}. We consider two cases for illustrative purposes,
choosing two polytropic indices, $n = 2.9$ (case $\textbf{N29}$) and  $n = 2.95$ (case
$\textbf{N295}$). According to~\cite{SunRuiSha18},
these choices describe stars with $\beta\sim 10^{-3}-10^{-2}$, which correspond to
$M\sim 10^4-10^5 M_\odot$, respectively. Their viscous timescales are computationally
realizable with the resources available to us. To match the rotation profiles in
\cite{ZinSteHaw05,ReiOttAbd13}, we choose the initial differential stellar rotation law as
\begin{equation}
  u^tu_\phi=\frac{ R^2_{eq}}{9}\,(\Omega-\Omega_c)\,,
  \label{eq:rotation_pro}
\end{equation}
where  $\Omega\equiv u^\phi/u^t$  is the angular velocity, and $\Omega_c$ is the angular
velocity along the rotating axis. For nearly Newtonian configurations, the above rotation
law reduces to
\begin{equation}
\label{eq:rotation_pro_newt}
\Omega \approx \frac{\Omega_c}{1 + \left({9\,\varpi}/{R_{\rm eq}}\right)^2}\,,
\,\,\,(\rm Newtonian)\,
\end{equation}
with $\varpi^2 =x^2 +y^2$. We set the ratio of the  polar to equatorial radius to 
$R_{\rm pol}/R_{\rm eq}=0.6$ and  the central density low enough
to guarantee dynamical radial stability. Both models have $T/|W| \approx 0.09$,
which is both secularly and dynamically stable to the $m=2$ mode, in contrast to the secularly
unstable models with $T/|W| = 0.227$ in~\cite{ZinSteHaw05, ReiOttAbd13}. 

The star is seeded with a dipole-like, dynamically weak magnetic field confined to the
stellar interior~(see left panels in Fig.~\ref{fig:density_B}). The field is generated
by the vector potential~\cite{SunPasRui17,SunRuiSha18},
\begin{align}
  A_\phi^{\rm int} = A_{ b}\,\varpi^2\,\text{max}
  (P-P_{\text{cut}}, 0)^{n_b}\,,
  \label{eq:A_int}
\end{align}
where $A_{ b}$, $P_{\rm cut}$, and $n_b$ are constants that determine the strength, the degree
of central condensation, and the confinement of the magnetic fields, respectively. Following
\cite{SunPasRui17,SunRuiSha18},  $P_{\rm cut}$ is set to $10^{-4}$ times the initial maximum value of the
pressure and $n_b$ to $1/8$. We then set the amplitude $A_b$ such that $\mathcal{M}/W\approx 2.0
\times 10^{-4}$ for the two cases. The choice of the field strength is based on a compromise between
insuring computationally achievable Alfv\'en and viscous damping timescales and preventing significant
departures from hydrostatic equilibrium triggered by a dynamically strong magnetic field.

We cover the computational grid with a tenuous atmosphere with constant density $\rho_{0,\,{\rm atm}}=10^{-10}\,
\rho_{0,\,\rm max}(0)$ following the standard setup in hydrodynamic and MHD simulations, where
$\rho_{0,\,\rm max}(0)$ is the maximum rest-mass density at $t = 0$. The key initial parameters are
summarized in Tables~\ref{tab:initial}, and~\ref{tab:initial_2}.

Before we start the numerical evolution of the above stars, we verify that MRI will be properly
captured in our system. Following~\cite{Gold:2013zma}, we check that: a) the magnetic field generated
by the vector potential in Eq.~\ref{eq:A_int} satisfies $\partial_{\varpi}\Omega<0$, a condition need
to trigger the instability. b) The  wavelength  of  the  fastest-growing mode $\lambda_{\rm MRI}$ should
be resolved by more than $10$ grid points~\cite{ShiLiuSha06,SanInuRTur04,ShiDolCha12}.  We plot the
quality factor $Q=\lambda_{\rm MRI}/dx$ on the equatorial plane, with $dx$ the local grid spacing.
As is showed in Fig.~\ref{fig:MRI-xy}, we resolved $\lambda_{\rm MRI}$ by more than $15$ grid points
in the bulk of our stellar models except for the ring-shaped region where the magnetic field flips direction.
c) The wavelength $\lambda_{\rm MRI}$ should fit in the star. Fig.~\ref{fig:MRI-xz}
shows $\lambda_{\rm MRI}$ along with the rest-mass density on the meridional plane. It is clear
that the wavelength  of the fastest-growing mode $\lambda_{\rm MRI}$ fits in the star. We conclude
therefore that the MRI will be resolved and operating during the evolution of our stellar models.
%
%
\section{Methods and Diagnostics}
\label{section:methods}
We solve Einstein's equations for the gravitational field coupled to the equations of MHD
for the matter and magnetic field. Numerically, we use the moving--grid adaptive mesh
refinement (AMR) Illinois GRMHD code embedded in the {{\tt
    Cactus}\footnote{http://www.cactuscode.org}/{\tt Carpet}\footnote{http://www.carpetcode.org}}
infrastructure. The code solves the Baumgarte--Shapiro--Shibata--Nakamura (BSSN) formulation
of the Einstein's equations~\cite{ShiNak95, BauSha98b} to evolve the metric and employs
moving puncture gauge conditions~\cite{BakCenCho05,CamLouMar05}. Here, we solve the equation
for the shift vector in first-order form (see e.g.~\cite{HinBuoBoy13,RuiHilBer10}) with the
shift parameter $\eta=0.66/M$ for~\textbf{N29} case and~$\eta=0.77/M$ for \textbf{N295} case,
with $M$ the ADM mass of the system. In solving the MHD equations, the code introduces a vector
potential (see Eqs. (8)-(9) in \cite{EtVpas12}) in order to maintain the divergence--free
condition of the magnetic field. The code solves the MHD equations in a conservative formulation
(see Eqs.(27)-(29) in~\cite{Etienne:2010ui}) using high-resolution shock-capturing methods
\cite{DueLiuSha05}. In addition, a generalized Lorenz gauge~\cite{EtVpas12,FarGolPas12} is used
to avoid the spurious magnetic fields between AMR levels due to numerical interpolations. As
the standard numerical tool, the Illinois GRMHD code has been extensively tested and has
performed multiple simulations, including merging magnetized compact objects and the collapsing
massive stars (see e.g.~\cite{EtVpas12,Paschalidis:2013jsa,SunPasRui17,KhaPasRui18,SunRuiSha18}
and references  therein for the suite of tests).
%
\begin{table}[t]
  \caption{\label{tab:resolution} Grid structure for all cases listed in Table~\ref{tab:initial}
    and~\ref{tab:initial_2}. The computational mesh consists of one set of j-nested AMR grids
    centered at the star, in which $j =1,...,6$ and equatorial symmetry is imposed. 
    The grid spacing for a given set of $j$-nested grids is denoted by
    $\Delta x_{\rm}$.  The half-side length of the outermost AMR boundary is given in the grid hierarchy with $j = 1$.}
\begin{ruledtabular}
\begin{tabular}{ccccc}
Case& $\Delta x_{\rm}$  & $\Delta x_{\rm} \left(M/10^4M_\odot\right){\rm km}$ & Grid hierarchy& \\
\hline
\textbf{N29}          &  $35.9M/2^{j-1}$ &$5.5\times 10^5 /2^{j-1}$ & $897M/2^{j-1}$ \\
\textbf{N295}            &  $31.2M/2^{j-1}$ &$4.8 \times 10^5 /2^{j-1}$ & $1000M/2^{j-1}$ \\ 
\end{tabular}
\end{ruledtabular}
\end{table}
%
%
\begin{figure*}[t]
  \includegraphics[width=0.493\textwidth]{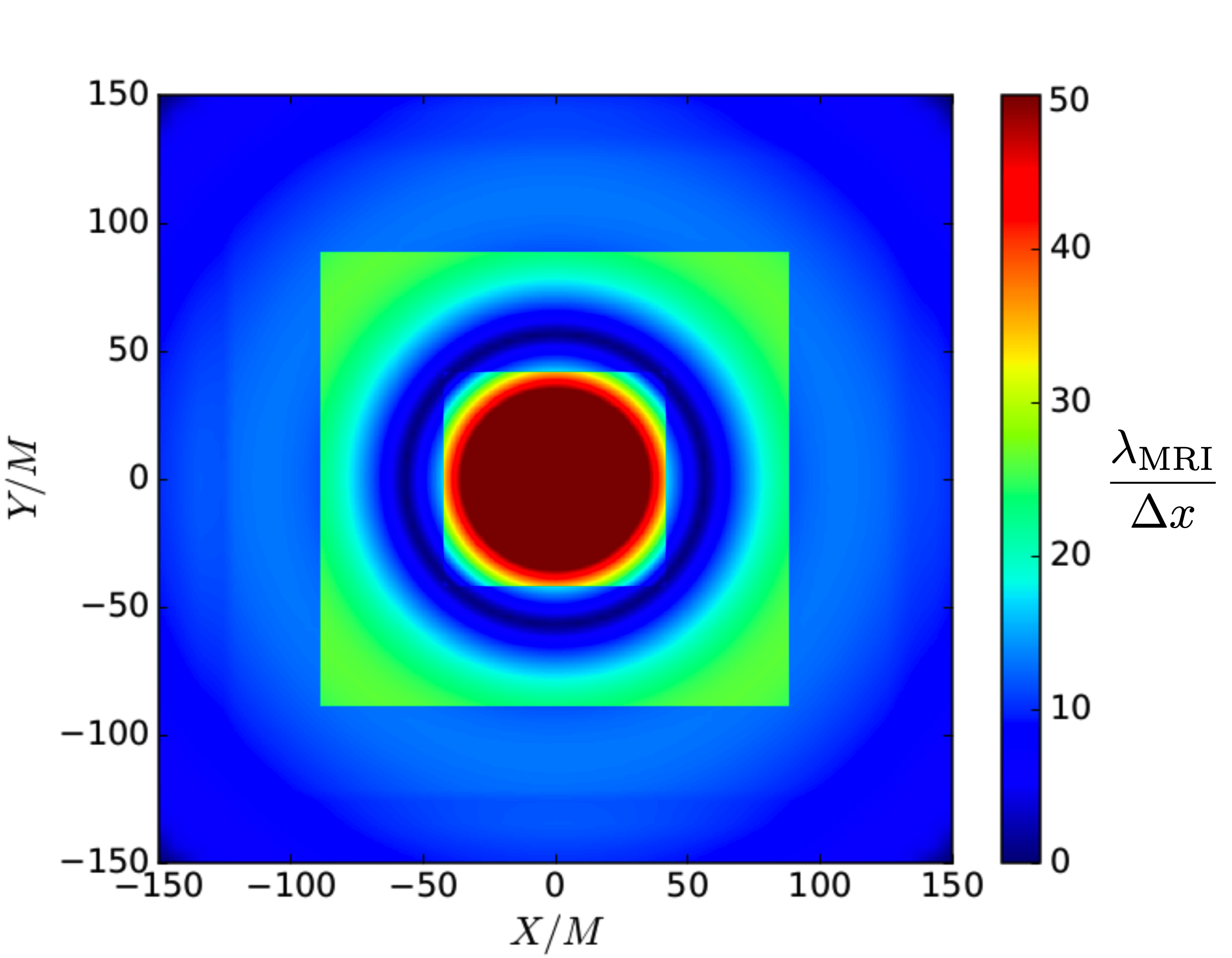}
  \includegraphics[width=0.493\textwidth]{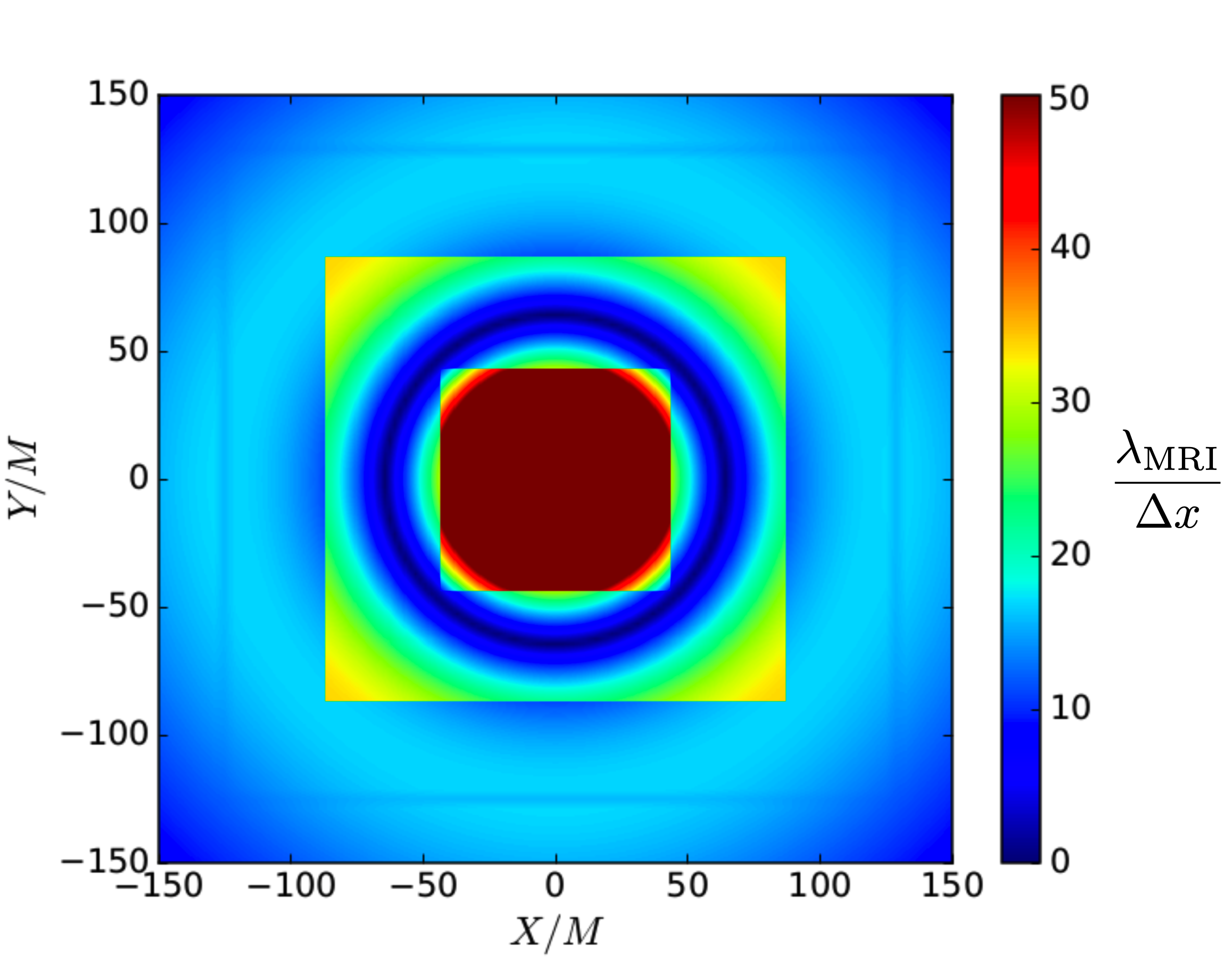}
  \caption{\label{fig:MRI-xy}
     Contours of the $\lambda_{\rm MRI}$ quality factor $q = \lambda_{\rm MRI}/\Delta x$  on
     the equatorial plane at $t=0$ for \textbf{N29} (left panel) and \textbf{N295} (right
     panel). Here $\Delta x$ is the grid spacing. The fastest growing MRI mode is resolved
     initially by~$\gtrsim 15$ gridpoints  in the bulk of the star. The blue ring regions indicate
     the flip of the direction of the magnetic field. Smaller boxes represent successively higher
     levels of refinement (see Table~\ref{tab:resolution}).}
\end{figure*}
To verify the reliability of our simulations, we monitor the normalized Hamiltonian and momentum
constraints employing Eqs.(40)-(43) in~\cite{Etienne:2007jg}, which remain below $\sim 1\%$ during
the evolution. We verify the conservation of the total mass $M_{\rm int}$ and total angular
momentum $J_{\rm int}$  computed using Eqs.(9)-(10) in~\cite{EtiLiuSha08}, which coincides with
the ADM mass and the ADM angular momentum at infinity, as well as the conservation of the
rest-mass $M_0$ described by Eq.(11) in~\cite{EtiLiuSha08}. We find that in the two cases listed
in Table~\ref{tab:initial} $M_{\rm int}$, $J_{\rm int}$, and $M_0$ deviate from their initial
values by only $\lesssim 0.5\%$ along the whole evolution.  We also monitor the conservation of the binding
energy $E_{\rm bind} = M_0-M_{\rm int}$, which may be decomposed according to 
%
%
%
\begin{figure*}[t]
  \includegraphics[width=0.96\textwidth]{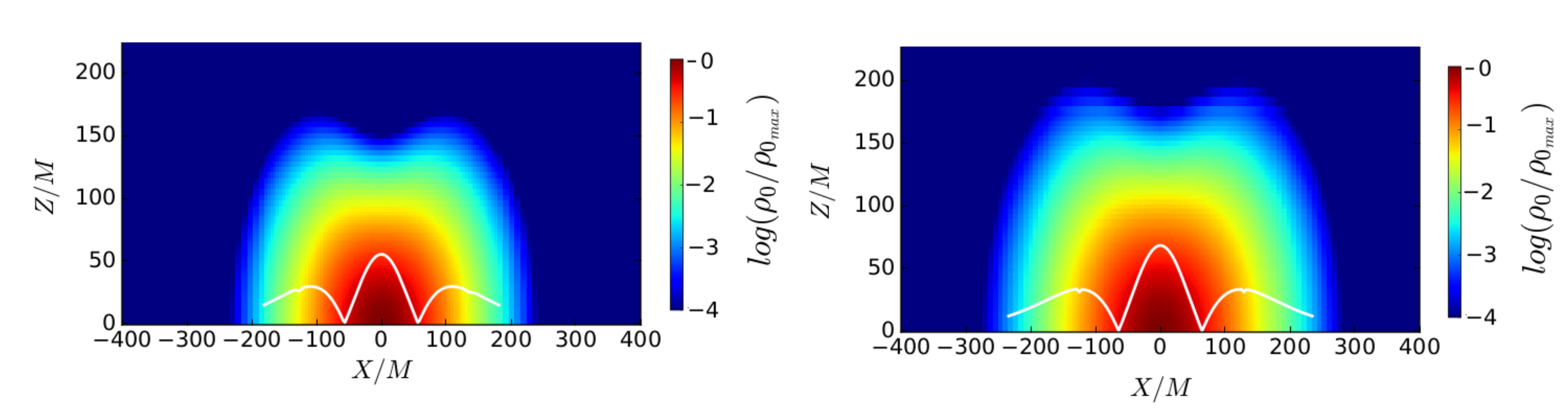}
  \caption{\label{fig:MRI-xz} Rest-mass density along with  $\lambda_{\rm MRI}$ (white line)
    on the meridional plane at $t=0$ for \textbf{N29} (left panel) and \textbf{N295} (right panel).
    The fastest growing MRI mode is mostly embedded in the star.}
\end{figure*}
%
%
\begin{eqnarray}
  -E_{\rm bind} &=&M_{\rm int} - M_0\nonumber\\ &=&
  E_{\rm int} + T + W +\mathcal{M}\,,
  \label{eq:E_bind}
\end{eqnarray}
where $E_{\rm int}$ is the internal (thermal) energy, $T$ is the rotational kinetic
energy, $\mathcal{M}$ is the magnetic energy, and $W$ is gravitational potential energy. These
quantities are calculated as follows:
\begin{equation}
\label{eq:Eint}
E_{\rm int} = \int_V (\rho_0\,\epsilon)\,d\mathcal{V}\,,
\end{equation}
\begin{equation}
\label{eq:T}
T = \int_V \frac{1}{2}\,\Omega\,T^0_{\rm fluid\,\phi}\,(u^0)^{-1}\,d\mathcal{V}\,,
\end{equation}
\begin{equation}
\label{eq:Eem}
\mathcal{M} = \int_V n_\mu\,n_\nu\,T^{\mu\nu}_{EM}\,(\alpha\,u^0)^{-1}\,
d\mathcal{V}\,,
\end{equation}
and
\begin{equation}
\label{eq:W}
W = M_{\rm ADM} - M_0 - E_{\rm int} - T - \mathcal{M}\,,
\end{equation}
where $d\mathcal{V} = \alpha\,u^0\,e^{6\phi} d^3 x$  is the proper volume
element with $\alpha$ the lapse function, and $\phi$ the conformal exponent
$\phi=\ln(\gamma)/12$, where $\gamma$ is the determinant of the three-metric $\gamma_{ij}$,
and  $\epsilon$ is the specific internal energy.  For the evolution, we use a $\Gamma$-law
EOS, $P=(\Gamma-1)\,\epsilon\,\rho_0$ and allow for shock heating. Here $\rho_0$ denotes the
rest-mass density. We set the adiabatic index $\Gamma$ to $\Gamma=1+1/n=1.345$ for~\textbf{N29}
and to $1.339$ for~\textbf{N295}.

To properly resolve the magnetic instabilities, in the two equilibrium stellar models in Table
\ref{tab:initial}, we use one set of mesh nested refinement boxes centered at the star. The
finest box has a grid spacing of {$\sim 30M/2^5\sim 1.4\times 10^4(M/10^4M_\odot)\rm km$},
and the star is resolved by more than 250 grid points across the diameter. The grid hierarchy for
each case is summarized in Table~\ref{tab:resolution}. We impose reflection symmetry across the
equatorial plane ($z=0$).
%
\section{Numerical Results}
\label{section:Results}
The basic dynamics and final outcome of our two cases listed in Table~\ref{tab:initial} are
similar. Redistribution of angular momentum via magnetic winding and a turbulent viscosity
triggered by magnetic instabilities in the bulk of the star induces the formation
of a massive and near uniformly rotating inner core wrapped by a low-density Keplerian-like
disk~(see right panels in~Fig.~\ref{fig:density_B}).

During the early phase of the stellar evolution, the dynamically weak magnetic field has a
negligible back-reaction on the fluid. The poloidal frozen-in magnetic field is advected
with the fluid and wound into a predominantly toroidal configuration. Differential rotation
in the bulk of the star stretches the field
lines, which in turn amplifies the toroidal component of the magnetic field (magnetic winding) and
increases the magnetic stresses. After about two rotation periods (see Fig.~\ref{fig:Omega_vs_R}),
back-reaction of the magnetic field is large enough to change the fluid motion and to redistribute
the angular momentum from the inner layers of the star to the outer regions~\cite{ShibataBook}.
This effect continues as long as the condition $B^l\,\partial_l\Omega\neq 0$ is maintained.
%
%
\begin{figure*}
\includegraphics[scale=0.18]{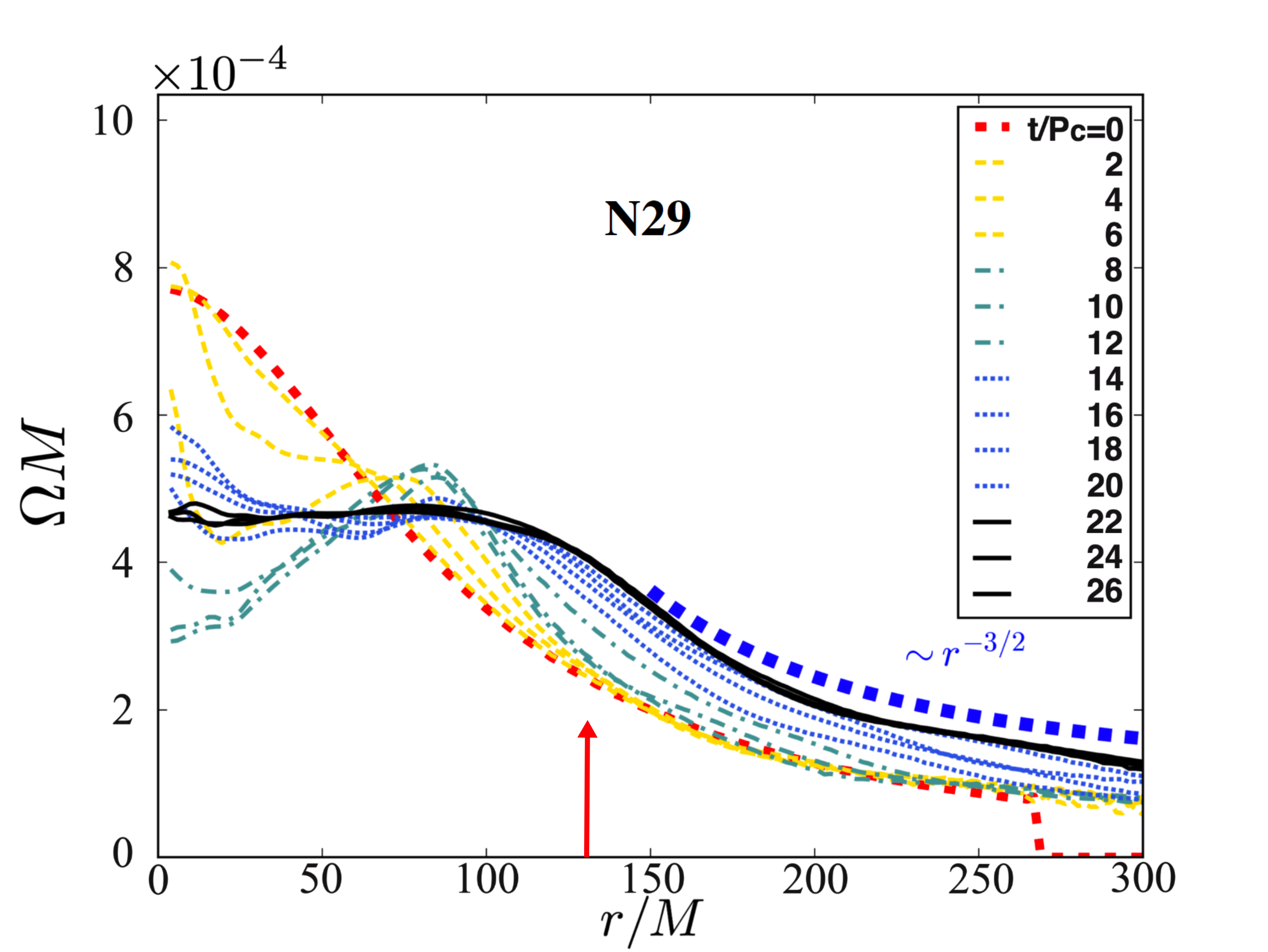}
\includegraphics[scale=0.18]{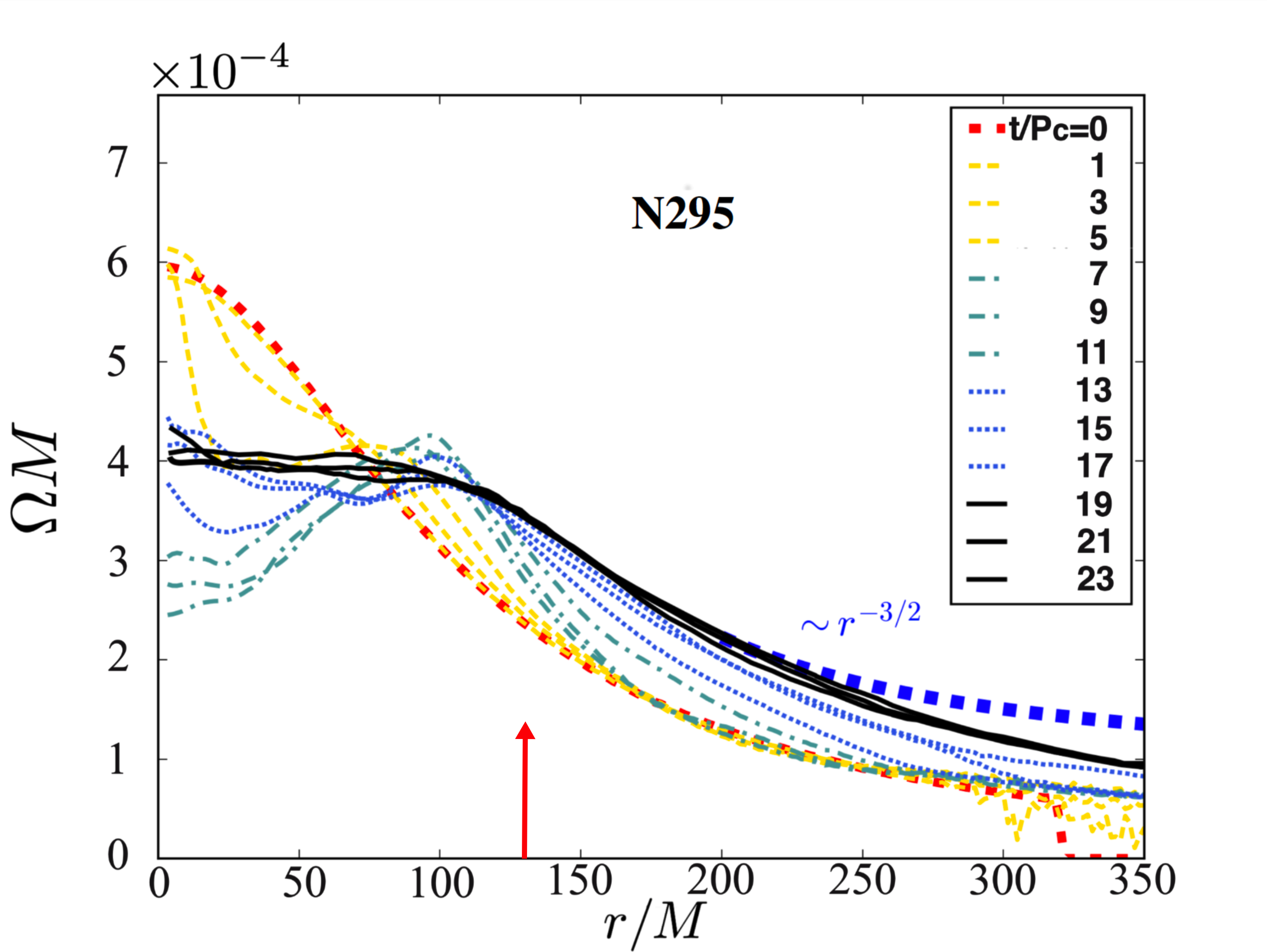}
\caption{\label{fig:Omega_vs_R} Angular velocity profile on the equatorial plane at multiples
  of initial central rotation period for case~\textbf{N29} (left panel) and \textbf{N295}
  (right panel). Curves are shown at equal time intervals in units of the initial period
  $P_{\rm c} = 8193M$ for \textbf{N29} and $P_{\rm c} = 10582M$ for \textbf{N295}. Here $M=0.049
  (M/10^4 M_{\odot})$s. The red dashed curve displays the initial differential rotation profile.
  The final profile is shown by the dark black line. The thick, blue dashed curve at large $r/M$
  shows the shape of a Keplerian profile. The spherical radius containing $1/2$ of the stellar
  rest-mass is denoted by the arrow.}
\end{figure*}

On the other hand, the timescale for the MRI can be estimated as
 \begin{equation}
   t_{\rm MRI} \sim 1/ \braket{\Omega} \sim 10^3 \left(
   \frac{\braket{\Omega}}{10^{-3} {\rm{\,rad \,s^{-1}} }}\right)^{-1} {\rm s}\,,
 \end{equation}
 where $\braket{\Omega} \sim 10^{-3} \rm{\,rad \,s^{-1}}$ is the average initial angular velocity.
 So, in the early phase of the stellar evolution,  the MRI is quickly triggered. This instability
 induces an effective viscosity that eventually induces an
 exponential growth of the magnetic energy (see Fig.~\ref{fig:Eem_vs_x}). Once the magnetic
 turbulence is fully developed, the angular momentum is redistributed by the turbulent viscosity
 from the inner (rapidly rotating) region to the outer  (slowly rotating) region. This effect
 continues as long as the condition $\partial_{\varpi}\Omega<0$ is satisfied.
 {Notice that at $t\sim 10\,P_c\sim t_{\rm A}$ inner layers of the star have
     an angular velocity that has a positive radial gradient (see Fig.~\ref{fig:Omega_vs_R}).
     So, at this time magnetic winding is mainly the mechanism that redistributes the
     angular momentum from the inner to the outer regions of the star.}
 
 Therefore, both magnetic winding and the MRI enhance the angular momentum redistribution.
 It causes the contraction of the inner stellar layers and the increase of central density,
 forming a massive core. We define the core as the stellar region with rest-mass density
 $\rho_0\geq 10^{-3}\,\rho_0^{\rm max}(0)$ where $\rho_0^{\rm max}(0)$ is the initial maximum
 rest-mass density. We estimate that, near to the end of the simulation, the radius of the inner
 core in the two stellar models considered here is roughly~$\sim 100M \sim 1.47\times 10^6 (M/10^4
 M_{\odot}) \rm km$~(see~right panels in Fig.~\ref{fig:density_B}). 
     After about twenty rotation periods, magnetic viscosity, which is at least partially developed,
     and magnetic stresses triggered by the MRI~(see Eq~\ref{eq:T_vis}) and magnetic winding redistribute
     the angular momentum and~bring the star to a new quasiequilibrium configuration:
     a nearly uniformly rotating central core + a Keplerian disk (see~\ref{fig:Omega_vs_R}).
  The new configuration then evolves on
 a secular timescale and remains in a quasi-stationary equilibrium until the termination of the simulation.
%
%
\begin{figure*}[t]
\includegraphics[scale=0.22]{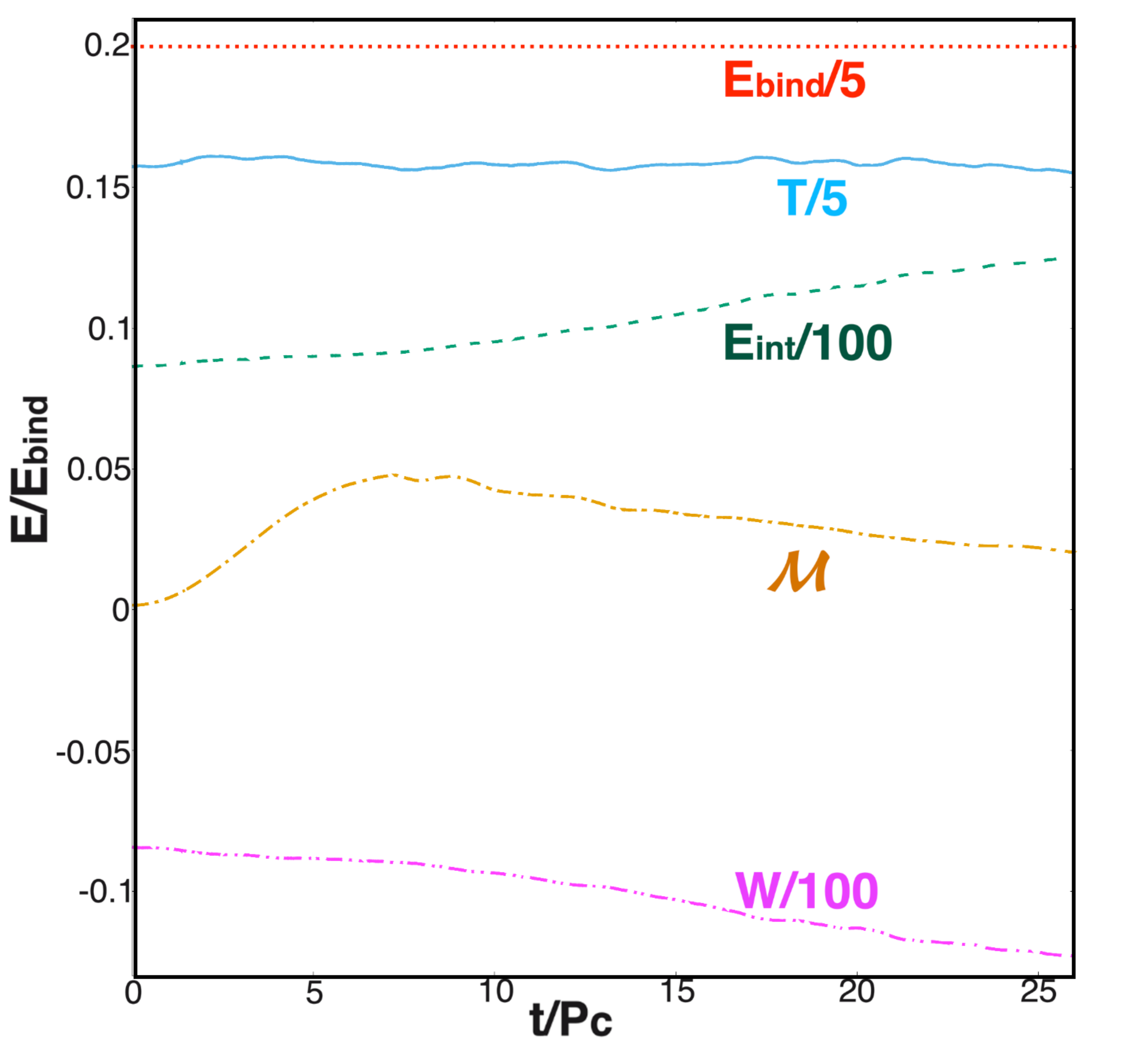}
\includegraphics[scale=0.22]{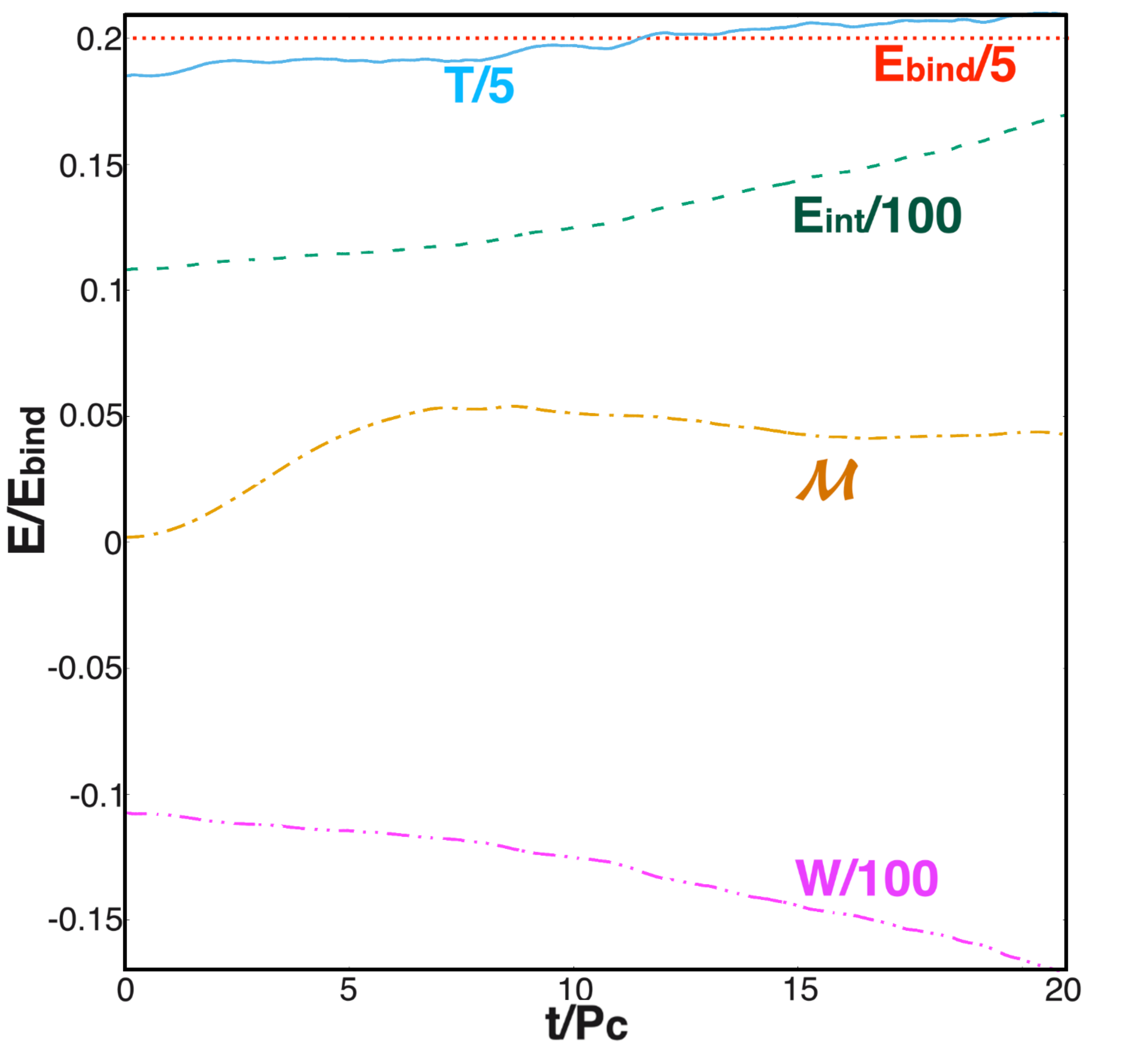}
\caption{\label{fig:Eem_vs_x} Evolution of the various energy components for~\textbf{N29}
  (left panel) and~\textbf{N295}~(right panel). All energies are normalized to the total binding
  energy at $t = 0$ (see Eq.~\ref{eq:E_bind}). The constituent energies are defined in Eqs.~\ref{eq:Eint}-~\ref{eq:W}.
  The binding energy $E_{\rm bind}$ should be conserved. Some quantities are normalized by
  an additional numerical factor (as indicated) to ease visualization.}
\end{figure*}

\paragraph*{\bf Time evolution of energies}
As previously stated, during the early phase of the stellar evolution, both magnetic winding and
the MRI enhance the magnetic energy $\mathcal{M}$. As shown in Fig.~\ref{fig:Eem_vs_x}, the magnitude
of $\mathcal{M}$ increases by roughly three orders of magnitude by $t\sim 6\,P_c$ where $P_c$
is the initial central rotation period, and then decreases due to damped oscillations.
The decrease of the magnetic energy is caused by turbulent viscosity triggered by the MRI and this
serves to heat the gas. After about $t\gtrsim\,20\,P_c$, the magnetic energy reaches a value of
$\sim 2.5\%$ of the binding energy in case
\textbf{N29} and $\sim 4.0\%$ in case \textbf{N295}. The internal energy $E_{\rm{int}}$ increases
gradually with time, indicating continuous heating. The major heating energy coincides with
the decrease of gravitational potential energy $W$ due to the contraction of
the stellar core (see Eq.~\ref{eq:T_th_2}). Note that, $\mathcal{M}$ is not the main heating source
because $\mathcal{M} \ll E_{\rm int}$ throughout the evolution. 

The rotational kinetic energy $T$, on the other hand, remains nearly constant throughout the
whole simulation in~\textbf{N29}, where we observe very small fractional changes. As shown in
Fig.~\ref{fig:Omega_vs_R}, after $t\gtrsim\, 20\,P_c$, the inner core remnant
spins at $\sim 60\%$ of the initial angular velocity of the star. In contrast, $T$ in~\textbf{N295}
increases by around $10\%$ of its initial value.
After $t\gtrsim\, 20\,P_c$, the inner core remnant spins at $\sim 66\%$ of the initial angular
velocity of the star. Notice that the gravitational potential energy $W$ decreases faster
in~\textbf{N295} than~\textbf{N29}. This behavior suggests that the increase of $T$ comes from
$W$, indicating that the stiffer the EOS, the more difficult it is to transfer potential energy. 
These results display great differences from previous studies of magnetic braking
of differential rotation in infinite cylinders and incompressible stars, where a substantial
fraction of $T$ is transferred to $\mathcal{M}$~\cite{Sha00,CooShaSte03,LiuSha03}. However,
the energy evolution of our models shares several similarities with the B1 model in the study
of magnetic braking and damping of neutron stars~\cite{DueLiuSha06}, where a non-hypermassive
and ``ultra-spinning" (angular momentum exceeding the maximum for uniform rotation at the same
rest mass) $\Gamma = 2$ polytrope with an effective turbulent viscosity were evolved. 

\section{Conclusions}
\label{section:Conclusions}
The growth of $m=2$ seed density perturbations and their fragmentation in massive stars with
a high degree of differential rotation undergoing collapse has been suggested as a viable
channel for binary black hole formation. Here we illustrate the likely fate of such massive
stars during their prior, stable, evolutionary lifetime. We build GR stellar models
described by a polytropic EOS with polytropic index $n=2.9$ and $n=2.95$, with the same
differential rotation law as in~\cite{ZinSteHaw05, ZinSteHaw06}.  We endowed
the stars with a dynamically weak, poloidal magnetic field confined to the stellar
interior. The ratio of initial magnetic energy to gravitational potential energy is
$\sim 2 \times 10^{-4}$.

We found that during the early phase of the evolution, magnetic winding and the
MRI greatly amplify the initial magnetic field. During that period, the magnetic
field is increased by $\sim 50$ times its initial value. At the
same time, angular momentum from the inner layers of the star is transported outward.
The inner layers then shrink and form a massive inner core in which the differential
rotation eventually (around twenty rotation periods) is fully damped. Therefore, 
prior to reaching the state of gravitational radial instability, magnetic braking and
turbulent magnetic viscosity drive a massive star to a new quasi-equilibrium state.
This state consists of a uniformly rotating central core surrounded by a low-density
Keplerian disk. These are not the matter or rotation profiles that will lead
to growth and fragmentation of $m=2$ density perturbations during catastrophic collapse.

We estimated various physical timescales that determine the stellar evolution of
the stellar models considered here. Using the initial parameters of our GR stellar
models, we found that the timescales obey a natural hierarchy
(see Eq.~\ref{eq:hierarchy}). This hierarchy shows that during the normal evolutionary
lifetime of a star, a sufficiently high degree of differential rotation required to trigger
$m=2$ perturbation growth and fragmentation leading
to binary BH formation during collapse will be suppressed. Our numerical results should
describe the fate of stars with characteristic masses
greater than several hundred solar masses, which are possible progenitors of binary
massive stellar-mass BHs and IMBHs that produce GW signals detectable by LVSC.

Given our numerical results, we conclude that the binary black hole formation channel
described in~\cite{ReiOttAbd13} is not likely, assuming that the differentially rotating
massive star that might undergo fragmentation during collapse experienced a  long
evolution phase in hydrostatic equilibrium before arriving at radial instability to
collapse. During such a phase, even a small initial magnetic field will amplify and
ultimately damp the differential rotation required to grow the $m=2$ seed perturbations
that trigger fission or fragmentation. However, if a high degree of differential
rotation  is resuscitated following some sudden catastrophic event, such as a massive
star binary collision and merger, followed by a sudden collapse, then it may still be
possible for the fragmentation-binary BH formation mechanism to be triggered. The likelihood
of such an event remains to be determined. 

{We remark that the values of the $\alpha_{\rm SS}$ parameter found numerically depend
strongly on the adopted resolution. As it has been shown in~\cite{Kiuchi:2017zzg,Hawley:2013lga,2011ApJ...738...84H} 
the higher the resolution, the higher the alpha parameter. Therefore, it is expected
that in higher order numerical simulations the viscosity timescale, the time over
which the turbulent viscosity damps differential rotation, will be shortened. But, magnetic
winding and the MRI still will drive a massive star to a new quasi-equilibrium state that
consists of a uniformly rotating central core surrounded by a low-density Keplerian disk.
Our basic conclusion during its lifetime therefore will remain unaffected, though the hierarchy
of timescales in Eq. \ref{eq:hierarchy} may change.

\begin{acknowledgments}
  We thank V. Paschalidis for useful discussions and the Illinois
  Relativity group REU team (K. Nelli) for assistance in creating
  Fig.~\ref{fig:density_B}. This work has been supported in part by
  National Science Foundation (NSF) Grant PHY-1602536 and PHY-1662211,
  and NASA Grant 80NSSC17K0070 at the University of
  Illinois at Urbana-Champaign. This work made use of the Extreme Science
  and Engineering Discovery
  Environment (XSEDE), which is supported by National Science Foundation
  grant number TG-MCA99S008. This research is part of
  the Blue Waters sustained-petascale computing project,
  which is supported by the National Science Foundation
  (awards OCI-0725070 and ACI-1238993) and the State of
  Illinois. Blue Waters is a joint effort of the University
  of Illinois at Urbana-Champaign and its National Center
  for Supercomputing Applications.
\end{acknowledgments}

\bibliographystyle{apsrev}
\bibliography{references}

\end{document}